\documentclass[%
 reprint,
 amsmath,amssymb,
 aps,
]{revtex4-1}

\usepackage{graphicx}
\usepackage{dcolumn}
\usepackage{bm}

\newcommand{\rel}{{\rm rel}}
\DeclareMathAlphabet{\mathpzc}{OT1}{pzc}{m}{it} 
\usepackage{braket,color,comment,amsmath}

\begin{document}

\preprint{APS/123-QED}

\title{Causal influence in linear response models}

\author{Andrea Auconi$^{1,2}$, Andrea Giansanti$^{2,3}$ and Edda Klipp$^1$}\email{edda.klipp@rz.hu-berlin.de}

\affiliation{$1$ Theoretische Biophysik, Humboldt-Universit\"at zu Berlin, Invalidenstraße 42, D-10115 Berlin, Germany\\
	$2$ Dipartimento di Fisica, Sapienza Universit\`a di Roma, Rome, Italy\\
	$3$ INFN, Sezione di Roma 1, Rome, Italy}

\date{\today}

\begin{abstract}
The intuition of causation is so fundamental that almost every research study in life sciences refers to this concept.
However a widely accepted formal definition of causal influence between observables is still missing.
In the framework of linear Langevin networks without feedbacks (linear response models) we developed a measure of causal influence based on a decomposition of information flows over time. We discuss its main properties and compare it with other information measures like the Transfer Entropy. Finally we outline some difficulties of the extension to a general definition of causal influence for complex systems.
\end{abstract}

\maketitle


\textit{"The causes of all the appearances in nature are the conditions under which they reliably emerge".}

Arthur Schopenhauer\cite{schopenhauer2012world}

\section{Introduction}

In the classical description of physical systems, observable objects are interacting in a symmetric way and "causation" means that a particular configuration at time $t$ is followed by (it \textit{causes}) a new configuration at time $t+\tau$ which is univocally determined by the laws of nature.

In the case of complex and living systems\cite{bialek2005physical,tkavcik2008information,kholodenko2006cell} we are rarely able to provide a full mechanical description because of the overwhelmingly broad range of time scales involved; moreover is often hard even to observe (and define the boundaries of) the objects taking part in these processes and accurately measure their properties.
This lack of knowledge leads us to conceive and represent the world \textit{in the practical sense} as if there were some intrinsically free observables, the signals or stimuli, which influence the behavior of other observables, the responses, through asymmetric causal interactions. The signals are "free" meaning that their dynamics is not influenced by other observables. As an example, the fluctuations of nutrients in the environment are usually modeled as Random and the cell responses as the consequent activation of biochemical signaling pathways.

We see that the intuition of causal influence between observables originates from a probabilistic description of nature.
When we say that a signal provokes a response we mean that the knowledge of the status of the signal $x(t)$ at time $t$ gives some information on the evolution $y(t+\tau)$ at time $t+\tau$ of the response which cannot be extracted by the knowledge of the response $y(t)$ at time $t$ itself, i.e. which is not already present in a redundant way in $y(t)$.

The causal influence is a measure of non-redundant information flow over time. It quantifies the average effect that the observable states of the signal have on the states of the response after a time period. We define it in the framework of information decomposition\cite{williams2010nonnegative} as the unique information that the signal has on the evolution of the response, i.e. the time-lagged Shannon's mutual information $I(x(t),y(t+\tau))$ minus the redundancy $R(x(t),y(t);y(t+\tau))$.
While the Shannon's mutual information is known, the redundancy has to be defined.

Some previously defined redundancy measures have been demonstrated to all have the same trivial and unintuitive form in Gaussian systems\cite{barrett2015exploration}: they take as redundancy the minimum of the information on the output ($y(t+\tau)$) given by the sources ($x(t)$ and $y(t)$) regardless of the mutual information that the sources share, $R_{nn}=\min[I(x(t),y(t+\tau)),I(y(t),y(t+\tau))]$.
We define instead the redundancy as a composition of the mutual information between the two sources $I_{xy}=I(x(t),y(t))$ and the total information that they give together on the output $I_{tot}=I(y(t+\tau),(x(t),y(t)))$:
\begin{eqnarray}\label{R}
R(\tau)=\frac{1}{2} \ln(\frac{e^{2(I_{xy}+I_{tot})}}{e^{2I_{xy}}+e^{2I_{tot}}-1}),
\end{eqnarray}

We developed this definition in the framework of linear Langevin networks without feedbacks (linear response models) because of their analytical tractability and intuitiveness. In section II.A we review the information processing properties of the bi-dimensional linear response model. It consists of a fluctuating signal described by a unidimensional Ornstein-Uhlenbeck process\cite{uhlenbeck1930theory,gillespie1996exact} that linearly influences the dynamics of a response variable. There we will study the properties of the resulting measure of causal influence $C_{x\rightarrow y}(\tau)\equiv I(x(t),y(t+\tau))-R(\tau)$ and compare it with the time-lagged mutual information $I(x(t),y(t+\tau))$ and with the transfer entropy $TE_{x \rightarrow y}=I(x(t),y(t+\tau)|y(t))$. Then in section II.B we extend the definition to the multidimensional case of directed acyclic graphs and discuss some simple examples.

Causation is naturally linked with an asymmetry in time, since the effects are defined to be successive to the causes. The arrow of time is not given \textit{a priori} but is empirically understood, therefore the time asymmetry which is proper of causality can't be derived from first principles but should be incorporated in the definition of causal influence itself (defining it only for $\tau>0$).

We note here that since the information measures are based on steady-state probabilities, a definition of causal influence based on information measures is appropriate just for such systems in which steady-state probability densities are defined, i.e. for ergodic stationary processes. 

Our definition of causal influence does not apply to networks with feedbacks and nonlinearities. Some intrinsic difficulties in the search for a general definition are discussed in the Discussion section.

\section{Results}
\subsection{Basic Linear Response Model}

\begin{figure}
	\begin{center}
		\includegraphics[scale=0.25]{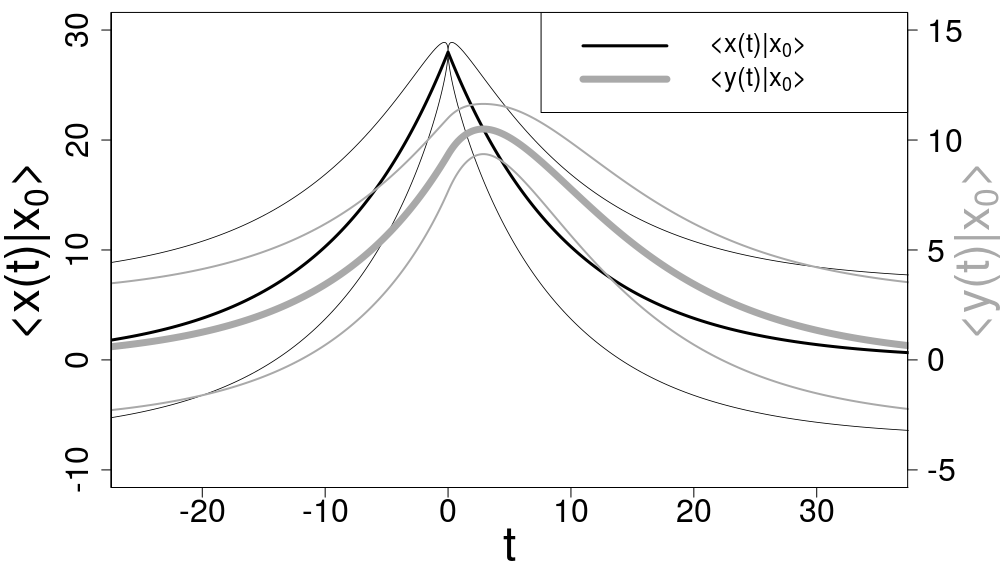}
		\caption{\label{due}Conditioned probability distributions over time. Given a particular condition (input) at time $t=0$, $x(0)\equiv x_0=28$, we plot the conditioned expectation values $\braket{y(t)|x(0)}$, $\braket{x(t)|x(0)}$ with the relative standard deviations $\sigma_{y(t)|x(0)}$, $\sigma_{x(t)|x(0)}$ (thinner lines) as a function of the shift $t$. The parameters are $\alpha=0.1$, $\beta=0.2$, $t_{rel}=10$, $D=10$.}
	\end{center}
\end{figure}

We define the Basic Linear Response Model (BLRM) with the two stochastic differential equations:
\begin{equation}\label{euno}
\begin{cases} 
\frac{dx}{dt} =-\frac{x}{t_\rel} + \sqrt{D}\,\, \Gamma(t) \\
\frac{dy}{dt} =\alpha x -\beta y
\end{cases} 
\end{equation} 
The $x(t)$ is a Ornstein-Uhlenbeck (OU) process\cite{uhlenbeck1930theory,gillespie1996exact}. It depends on the realization of the uncorrelated Gaussian noise $\Gamma$, which is defined as the $dt\rightarrow0$ limit of the normal random variable $\mathcal{N}(0,\frac{1}{dt})$.
The OU process describes fluctuations around zero mean with an autocorrelation function that decays exponentially, $\braket{x(t)x(t+\tau)}=\sigma_x^2 \exp\left(-\tau/t_\rel\right)$, where $t_\rel$ is the relaxation time, and whose amplitude is proportional to the square root of the diffusion coefficient $D$, $\sigma_x^2=D \frac{t_\rel}{2}$.
One easily shows that the probability density of $x(t+t')$ conditioned on $x(t)$ for generic (positive or negative) shifts $t'$ is given by the Gaussian distribution:
\begin{equation}\label{etre}
P(x(t+t')|x(t))=\mathpzc{G}\left[x(t)e^{-\frac{|t'|}{t_\rel}},
\sigma_{x}^2(1-e^{-\frac{2|t'|}{t_\rel}})\right],
\end{equation}
Equation (\ref{due}) describes a linear response to the OU process ($\alpha\neq0$ and $\beta>0$ are constants), whose formal solution is:
\begin{equation}\label{equattro}
y(t+\tau)=y(t)e^{-\beta \tau} + \alpha \int_{0}^{\tau}dt'x(t+t')e^{-\beta (\tau-t')},
\end{equation}

The BLRM is the simplest continuous-time dynamical system with evident causal influence: the dynamics of the variable $y$ is driven by the position of the variable $x$ which fluctuates around a mean value. The causal network is simply $x \rightarrow y$ and from now on we will call $x$ the signal and $y$ the response.

\begin{figure}
	\begin{center}
		\includegraphics[scale=0.25]{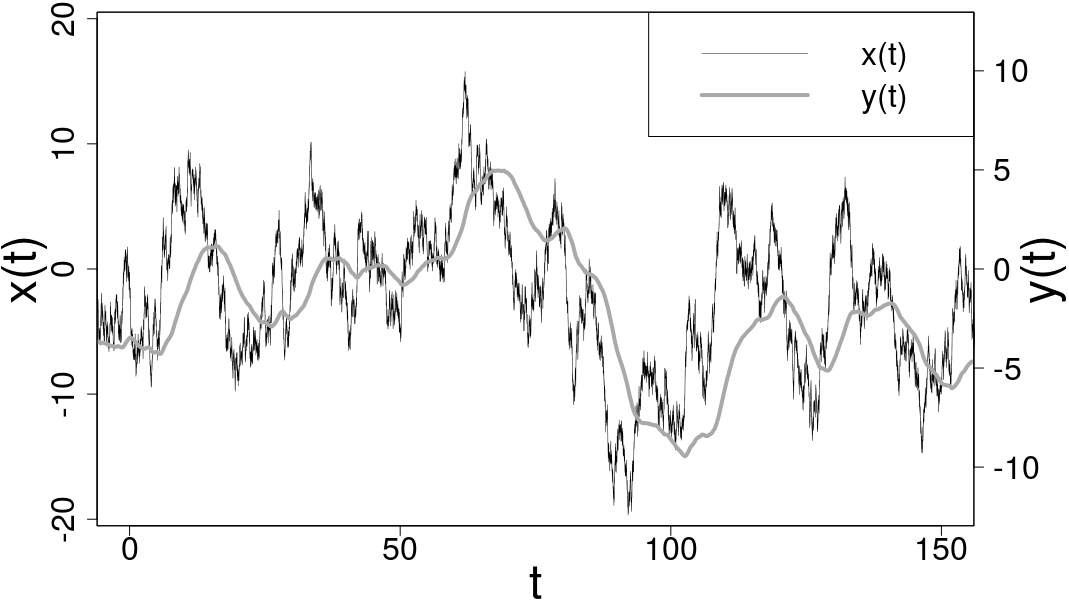}
		\caption{\label{uno}Stochastic dynamics of the Basic Linear Response Model. The parameters are $\alpha=0.1$, $\beta=0.2$, $t_{rel}=10$ and $D=10$. All graphs are produced using R\cite{team2013r}.}
	\end{center}
\end{figure}

We see (Fig.\ref{uno}) that the response is not following the signal in its "smallest" fluctuations but is rather integrating it on a time scale defined by the parameters $\beta$ and $t_{rel}$ as we will show.

Since the BLRM is a stationary process, the mean of the derivative of the variables products is vanishing $\braket{\frac{d(y^2)}{dt}}=\braket{\frac{d(xy)}{dt}}=0$, and the amplitude of the response fluctuations (driven by the signal) $\sigma_y$ results to be proportional to $\alpha$, $\sigma_y^2=\frac{\alpha^2 t_{rel}}{\beta(\beta t_{rel}+1)}\sigma_x^2$.

The expectation value of the response $y$ at time $t\pm\tau$, ($\tau>0$) conditioned to the knowledge of the signal $x$ at time $t$ (Fig.\ref{due}) is found using eq.\ref{etre}:
\begin{equation}\label{ecinque}
\braket{y(t-\tau)|x(t)}=x(t)\frac{\alpha t_{rel}}{\beta t_{rel}+1}e^{-\frac{\tau}{t_\rel}},
\end{equation}
\begin{equation}\label{esei}
\braket{y(t+\tau)|x(t)}=x(t)\frac{\alpha t_{rel}}{\beta t_{rel}-1}(e^{-\frac{\tau}{t_\rel}}-\frac{2e^{-\beta \tau}}{\beta t_{rel}+1}),
\end{equation}

The Gaussianity of the OU process implies the Gaussianity of the BLRM\cite{risken1984fokker}, so that the mutual information\cite{cover2012elements} that $x(t)$ and $y(t+t')$ share is just a function of their correlation $C(x(t),y(t+t'))$: 

\begin{equation}\label{esette}
I(x(t),y(t+t'))=\ln(\frac{\sigma_{y}}{\sigma_{y(t+t')|x(t)}})=-\frac{1}{2}\ln(1-C^2(x(t),y(t+t'))),
\end{equation}

The interpretation of the mutual information as the reduction in uncertainty of the variable $y(t+t')$ given the knowledge of the variable $x(t)$ is clearly seen in eq.(\ref{esette}). We note that $\sigma_{y(t+t')|x(t)}$ is independent of the particular condition $x(t)$.

Thanks to the fluctuation-dissipation theorem\cite{marconi2008fluctuation} we can also interpret the mutual information as a function of the signal-to-noise ratio $SNR$ which is the average square deviation from the mean of the estimate of $y(t+t')$ provoked by the knowledge of $x(t)$ divided by the uncertainty that remains $\sigma_{y(t+t')|x(t)}$:
\begin{equation}\label{nove}
I(x(t),y(t+t'))=\frac{1}{2}\ln(1+SNR),
\end{equation}
\begin{equation}\label{nove_b}
SNR=\frac{(\frac{\partial\braket{y(t+t')|x(t)}}{\partial x(t)})^2\sigma_{x}^2}{\sigma_{y(t+t')|x(t)}^2},
\end{equation}

The time shift which gives the optimal information transmission is given by:
\begin{equation}\label{dieci}
\tau_{opt}=\frac{t_{rel}}{\beta t_{rel}-1} \ln(\frac{2\beta t_{rel}}{\beta t_{rel}+1}),
\end{equation}
which is always positive as we expect since is the response $y$ that is following the fluctuations of the signal $x$ producing a natural delay in the correlation\cite{nemenman2012gain}.

The mutual information corresponding to the optimal time shift is:
\begin{equation}\label{undici}
I^{opt}=I(x(t),y(t+\tau_{opt}))=-\frac{1}{2} \ln(1-2(\frac{\beta t_{rel}+1}{2\beta t_{rel}})^\frac{\beta t_{rel}+1}{\beta t_{rel}-1}),
\end{equation}
which depends just on the product $\beta t_{rel}$ that we understand as the ratio of the two time scales of the model: the relaxation time of the signal fluctuations $t_{rel}$ and the response time of the system for a deterministic input $\frac{1}{\beta}$. The limits of high and low information are respectively $I^{opt}(\beta t_{rel}\rightarrow\infty)=\frac{ln(\beta t_{rel})}{2}$ and $I^{opt}(\beta t_{rel}\rightarrow 0)=2\beta t_{rel}$.

The mutual information is a fundamental characterization of the system, and in general one can use this measure to infer causal relations in network reconstruction\cite{kirst2016dynamic}. Still, it is not a measure of causal influence. The time-lagged mutual information is greater than zero for both cases $I(x(t),y(t+\tau))$ and $I(y(t),x(t+\tau))$, meaning that the response at time $t$ gives some information on the signal at successive times $t+\tau$ as we expect since the signal has autocorrelation. Nevertheless the response is not influencing the dynamics of the signal, therefore a measure of causal influence should be zero in this case. 

We could solve this problem considering instead the conditional mutual information, the so called Transfer Entropy\cite{schreiber2000measuring}(equivalent to the Granger Causality\cite{granger1969investigating} in linear systems), that is the additional amount of information that one gets on $y(t+\tau)$ upon knowledge of $x(t)$ when $y(t)$ was already known. For the BLRM it is given by:
\begin{widetext}
\begin{eqnarray}\label{dodici}
&  TE_{x \rightarrow y}= I(x(t),y(t+\tau)|y(t))=\ln(\frac{\sigma_{y(t+\tau)|y(t)}}{\sigma_{y(t+\tau)|x(t),y(t)}})=\\
&=  \frac{1}{2} \ln(1+\frac{\beta t_{rel} (e^{-\frac{\tau}{t_c}}-e^{-\beta \tau})^2}{(1-\beta t_{rel})^2-e^{-2\beta \tau}(1+\beta t_{rel})+e^{-(\beta +\frac{1}{t_{rel}})\tau}4\beta t_{rel}-e^{-\frac{2\tau}{t_{rel}}}\beta t_{rel}(1+\beta t_{rel})}), \nonumber 
\end{eqnarray}
\begin{eqnarray}\label{tredici}
TE_{y \rightarrow x}= I(y(t),x(t+\tau)|x(t))=\ln(\frac{\sigma_{x(t+\tau)|x(t)}}{\sigma_{x(t+\tau)|y(t),x(t)}})=0,
\end{eqnarray}
where $\sigma_{y(t+\tau)|x(t),y(t)}$ is the standard deviation of $y(t+\tau)$ conditioned to the knowledge of $x(t)$ and $y(t)$.
\end{widetext}

Here we have $TE_{y \rightarrow x}=0$ according to the fact that the dynamics of the signal is independent of the response.
$TE_{x \rightarrow y}$ is always positive instead and diverges for $\tau\rightarrow 0$ because the knowledge of both $x(t)$ and $y(t)$ synergistically provides  information on $y(t+\tau)$ since for small $\tau$ the $y_{t+\tau}$ map is quasi deterministic in $x_t$ and $y_t$: $y(t+\tau)-y(t)=\tau(\alpha x(t)-\beta y(t)) + o(\tau^2)$. This feature we would call determinism, while the causal influence is rather the visible (macroscopic) effect of the causation of the evolution of the response by the signal which is obtained gradually over time after the "cause" $x(t)$. It is the information on the evolution of the response $y(t+\tau)$ that we get looking at just the signal $x(t)$, $I(x(t),y(t+\tau))$, minus that part of information which is already present in a redundant way also in the response $y(t)$.

Following Barrett\cite{barrett2015exploration}, we define a decomposition of the information that $x(t)$ and $y(t)$ give on the evolution of the response $y(t+\tau)$:
\begin{eqnarray}\label{quattordici}
I(y(t+\tau),(x(t),y(t)))=R+U_x+U_y+S,
\end{eqnarray}

where $R$ is the redundancy, $U_x$ and $U_y$ are the unique information contributions respectively of $x(t)$ and $y(t)$ alone, and the synergy $S$ is defined as the information that one gets in addition when considering simultaneously both $x(t)$ and $y(t)$. People are trying to define the information decomposition in a way that the unique contribution from the signal $U_x(\tau)$ can be interpreted as a measure of the actual information flowing from the signal to the response over time\cite{james2016information}, that is what we call causal influence. Since the input-output mutual information is decomposed in $I(x(t),y(t+\tau))=R+U_x$ and the Transfer Entropy in $TE_{x \rightarrow y}=U_x+S$, in order to specify the decomposition it is sufficient to give a definition of the redundancy $R$.

\begin{figure}
	\begin{center}
		\includegraphics[scale=0.25]{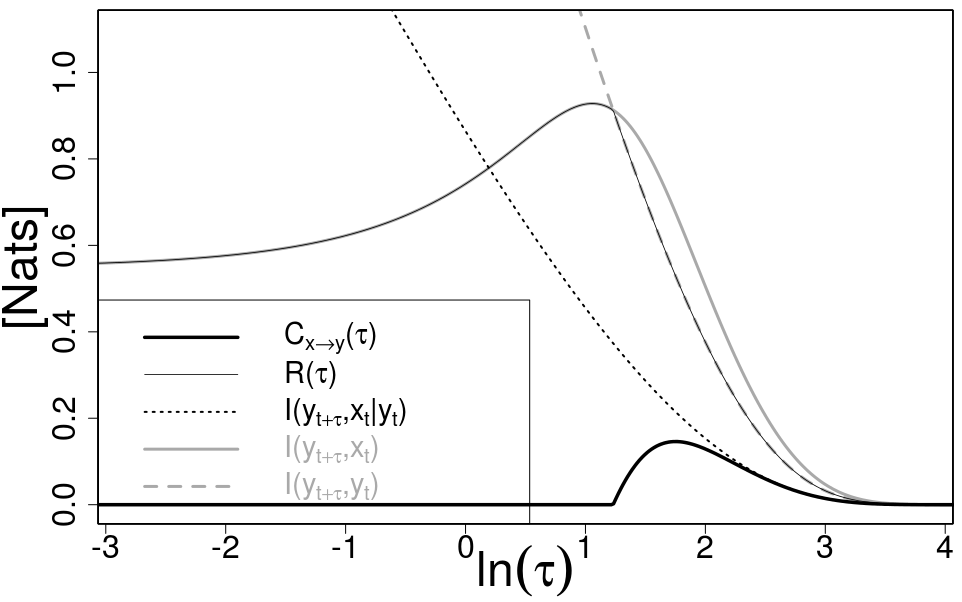}
		\caption{\label{tre}Nonnegative decomposition of Williams and Beer. The information measures are measured in natural units $Nats=\frac{bits}{\ln 2}$. The parameters are $\beta=0.2$, $t_{rel}=10$.}
	\end{center}
\end{figure}

In Gaussian systems\cite{barrett2015exploration} the previously defined measures of redundancy converge to the nonnegative decomposition of Williams and Beer\cite{williams2010nonnegative} which takes as redundancy the minimum value between $I(x(t),y(t+\tau))$ and $I(y(t),y(t+\tau))$, regardless of the information shared between the two sources $I(x(t),y(t))$, $R_{nn}=\min[I(x(t),y(t+\tau)),I(y(t),y(t+\tau))]$. With this definition (Fig.\ref{tre}) the unique information $U_x$ is zero until $I(x(t),y(t+\tau))$ is smaller than $I(y(t),y(t+\tau))$, suggesting the existence of an unintuitive activation time for causality. Moreover, let's consider the time shift $\tau=\tau_e$ for which $I(x(t),y(t+\tau_e))=I(y(t),y(t+\tau_e))$. According to the nonnegative decomposition, the mutual information should be totally redundant. However, redundancy does not mean that the two sources give the same information here since the estimate of $y(t+\tau)$ given $x(t)$ is in general different from the estimate of $y(t+\tau)$ given $y(t)$, while only the (reduction in) uncertainty of the estimates is the same; the similarity of these two estimates is given by $I(x(t),y(t))$ and the nonnegative definition of redundancy is not explicitly dependent on this.

We define instead the redundancy as a composition of the information shared between the two variables $x(t)$ and $y(t)$, $I_{xy}=I(x(t),y(t))$, and the total information that they share with $y(t+\tau)$, $I_{tot}=I(y(t+\tau),(x(t),y(t)))$  (eq.\ref{R} in the introduction):
\begin{eqnarray}
R(\tau)=\frac{1}{2} \ln(\frac{e^{2(I_{xy}+I_{tot})}}{e^{2I_{xy}}+e^{2I_{tot}}-1}) \nonumber
\end{eqnarray}

\begin{figure}
	\begin{center}
		\includegraphics[scale=0.25]{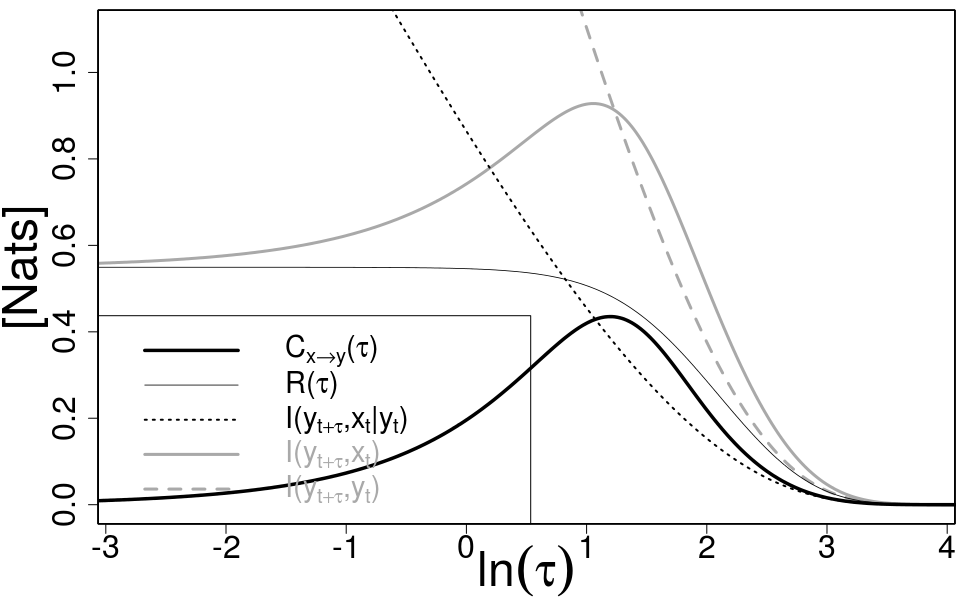}
		\caption{\label{quattro}Linear decomposition $x\longrightarrow y$. In thick black is the unique information that $x(t)$ gives on $y(t+\tau)$, that is our measure of causal influence $C_{x\rightarrow y}(\tau)$. The parameters are $\beta=0.2$, $t_{rel}=10$. The legend is the same as in Fig.\ref{quattro}.}
	\end{center}
\end{figure}

This definition is inspired by the logic of a linear Markov chain, i.e. we use the formula of the information shared between the variables $A$ and $C$ in the \textit{static} Gaussian linear network $A \rightarrow B \rightarrow C$. In other words, we define the redundancy as the information that $x(t)$ has on $y(t+\tau)$ "passing through" $y(t)$. The redundant information that the two variables $x(t)$ and $y(t)$ have both on the third variable $y(t+\tau)$ is a fraction of the mutual information that they share $I_{xy}$, and that fraction is determined by the total information that they have together on $y(t+\tau)$, $I_{tot}$. The redundancy measure is symmetric in the two sources (causes) $x(t)$ and $y(t)$, and also symmetric in $I_{xy}$ and $I_{tot}$. We call this information decomposition the Linear decomposition (Fig.\ref{quattro}) and we suggest that the proposed definition of redundancy (eq.\ref{R}) could be a linear approximation of a more general definition to be found.

\begin{figure}
	\begin{center}
		\includegraphics[scale=0.25]{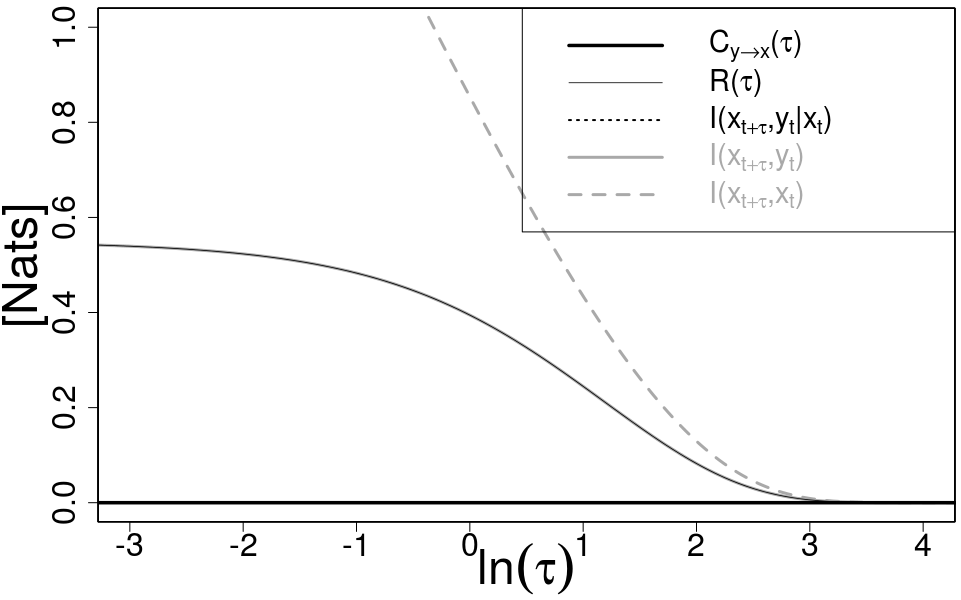}
		\caption{\label{cinque}Linear decomposition $y\longrightarrow x$. The redundant information is equal to the mutual information meaning that there's no causal influence. The parameters are $\beta=0.2$, $t_{rel}=10$.}
	\end{center}
\end{figure}

The unique information $U_x$ that results from eq.(\ref{R}), $U_x=I(x(t),y(t+\tau))-R(\tau)$, we claim to be a measure of causal influence for the LRM:
\begin{eqnarray}\label{Definizione}
C_{x\rightarrow y}(\tau)\equiv I(x(t),y(t+\tau))-R(\tau),
\end{eqnarray}
We define the causal influence $C_{x\rightarrow y}(\tau)$ only for positive $\tau\geq0$ to include the empirical knowledge that the effects are always seen after the causes. $C_{x\rightarrow y}(\tau)$ is a measure of information, therefore is measured in natural units $Nats$.

Given the realization of the OU process $x$ for a sufficiently long time we can determine with any precision the position of the $y(t+\tau)$, and it`s fair to say that this value is totally caused by the sequence of $x$ at previous times. This fact is mirrored in the divergence of the Transfer Entropy for small time shifts $\tau\rightarrow 0$. Nevertheless (in the common language) we usually consider as causes the single observable facts ($x(t)$ and $y(t)$ in the BLRM), and as the effect a successive observable fact ($y(t+\tau)$), and we wish to quantify the relative strength of these causes in giving the effect.
The causal influence $C_{x\rightarrow y}(\tau)$ that the signal has on the response over time is $0$ for the time shift $\tau=0$ and increases with $\tau$ (linearly for small $\tau$) meaning that we get the effect of causality (that is the causal influence) gradually over time after the cause $x(t)$. For very long time intervals $\tau$ after the cause we cannot see anymore the effect of the distant past and the causal influence goes to $0$. The time shift at which the causal influence peaks $\tau_{res}$ is the response time of the system in the probabilistic sense and is slightly different from the maximum correlation time $\tau_{opt}$. In general $\tau_{res}>\tau_{opt}$.
We note that, as it should be, we get zero causal influence of the response $y$ on the signal $x$ (fig.\ref{cinque}). This is because the information $I(y(t),x(t+\tau))$ that the response has on the evolution of the signal is gained necessarily via the two steps $y(t) \rightarrow x(t)$ and $x(t) \rightarrow x(t+\tau)$ due to the asymmetry of the interaction and therefore is equal to the redundancy $R(x(t),y(t);x(t+\tau))$.

\begin{figure}
	\begin{center}
		\includegraphics[scale=0.25]{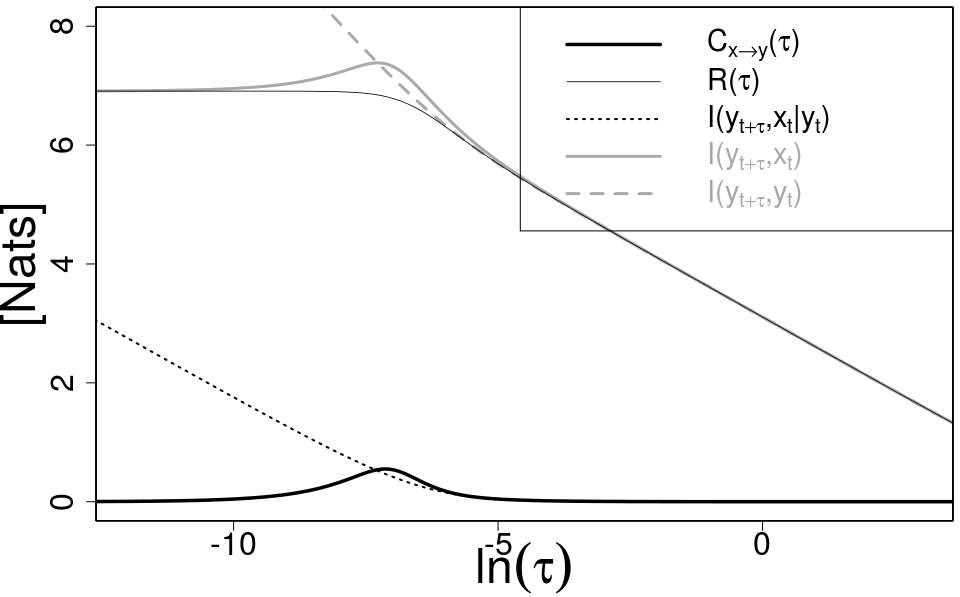}
		\caption{\label{A}Linear decomposition $x\longrightarrow y$. High information scenario: $\beta=1000$, $t_{rel}=1000$.}
	\end{center}
\end{figure}

\begin{figure}
	\begin{center}
		\includegraphics[scale=0.25]{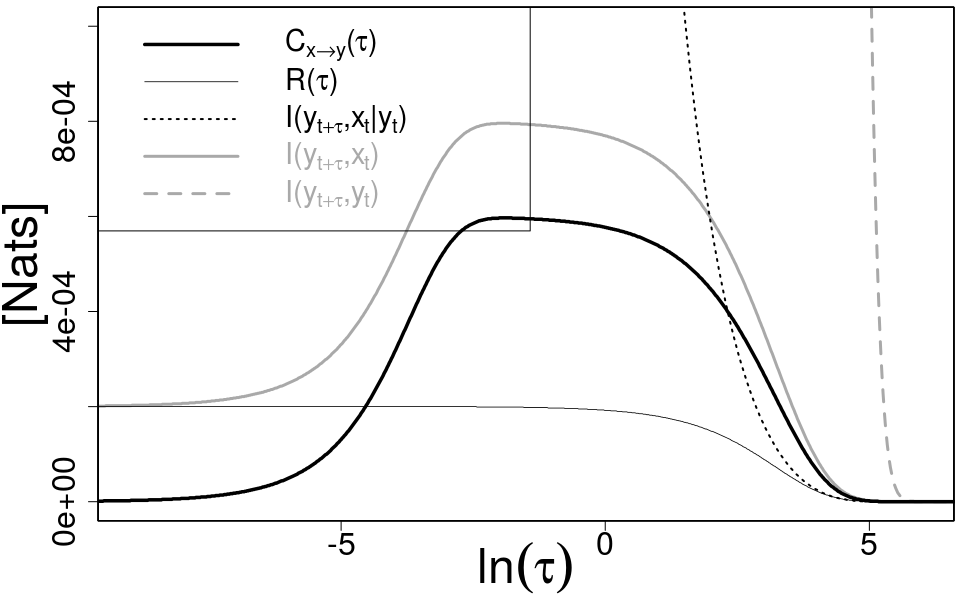}
		\caption{\label{B}Linear decomposition $x\longrightarrow y$. Low information scenario: $\beta=0.02$, $t_{rel}=0.02$.}
	\end{center}
\end{figure}

To understand the behavior of the causal influence in the BLRM as a function of the parameters we study the limits of high and low information (fig.\ref{A}-\ref{B}). When $\beta t_{rel}>>1$ the mutual information is high and increases with $\beta t_{rel}$. The peak of the causal influence also increases but only up to a limit of around $\lim_{\beta t_{rel}\rightarrow \infty} \max_\tau C_{x\rightarrow y}\approx 0.55 ~ Nats$. The position of the peak depends on $\beta$: with higher $\beta$ the response is faster and the effect of causality is seen earlier. When $\beta$ is fixed, increasing $t_{rel}$ gives always an increase in the mutual information because the slow down of the dynamics of the signal lets the response follow the microscopical structure of the signal with more precision, but at the same time the response is moved slowly (in units of his standard deviation) by the signal, these two effects asymptotically compensating and the causal influence staying around $\approx 0.55 ~ Nats$. This limit we call the causation capacity of the BLRM.
In the case of low information $\beta t_{rel}<<1$ the peak of the causal influence is close to $75\%$ of the peak of the mutual information because $\frac{I^{opt}}{I_{xy}}\rightarrow 4$ for $\beta t_{rel}\rightarrow 0$. The signal has fast-decaying autocorrelation, the response is slowly integrating (keeping the memory of) it and therefore most of the small amount of time-lagged mutual information on the response is causal influence.

Finally we note that the synergy $S=TE_{x \rightarrow y}-C_{x \rightarrow y}$ is negative when the causal influence is greater than the transfer entropy, and this is always the case for long delays $\tau$. This means that part of the "same" information that $x(t)$ and $y(t)$ give on $y(t+\tau)$ is considered as causal influence and not redundancy.

\subsection{Multidimensional case: networks without feedbacks}

We can extend the causal influence measure for interactions within linear Langevin networks without feedbacks.
Let us define the network of direct influences as the one that has directed links for all the combination of variables (nodes in the network) $(i\rightarrow j)$ for which the variable $i$ appears in the equation for the dynamics of the variable $j$.
The network of the causal influence is not coincident with the network of direct influences because we also have to consider as causal all the indirect influences.
Let us define the parents $P_{x}$ of a node $x$ as the set of all nodes in the network of direct influences that are able to reach $x$ with directed paths. We expect all the parents $P_{x}$ to have causal influence on $x$, in general with different intensities and time scales.
Similarly we define the parents $P_{xy}$ of two nodes $x$ and $y$ as the set of all nodes in the network of direct influences (excluding $x$ and $y$ themselves) that are able to reach $x$, $y$ or both nodes with directed paths.

Then we generalize the definition of causal influence adding the condition of the knowledge of the state of the parents $P_{xy}(t)$ at time $t$ to all the probability measures:
\begin{widetext}
\begin{eqnarray}\label{sedici}
C_{x\rightarrow y}(\tau)=I(x(t),y(t+\tau)|P_{xy}(t))-R(x(t),y(t);y(t+\tau)|P_{xy}(t))  ,
\end{eqnarray}
\end{widetext}

where $R(x(t),y(t);y(t+\tau)|P_{xy}(t))$ is defined as in eq.(\ref{R}) but with all the information measures conditioned to the knowledge of the parents $P_{xy}(t)$ at time $t$.

For simplicity we consider a network of three nodes without feedbacks, the so called Feed-forward loop:
\begin{eqnarray}\label{trio}
\begin{cases} 
\frac{dz}{dt} =-\frac{z}{t_\rel} + \sqrt{D_z}\,\, \Gamma_z(t) \\
\frac{dx}{dt} =\alpha_x z -\beta_x x + \sqrt{D_x}\,\, \Gamma_x(t) \\
\frac{dy}{dt} =\alpha_y z -\beta_y y + \gamma x  +\sqrt{D_y}\,\, \Gamma_y(t) 
\end{cases} 
\end{eqnarray}

  \begin{center}
   \includegraphics[scale=0.25]{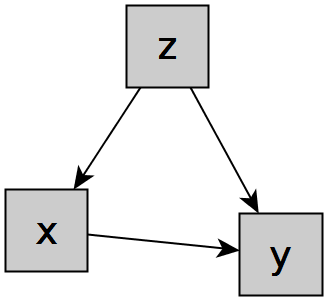}
  \end{center}    

When $\gamma=0$, the variable $x$ is not a parent of $y$ and therefore it should have no causal influence on it. Still, $x$ and $y$ can be highly correlated due to the common parent $z$. Applying the above definition we analytically calculated the causal influence of $x$ on $y$ and it resulted to be zero (see APPENDIX A), $C_{x\rightarrow y}(\tau)=I(x(t),y(t+\tau)|z(t))-R(x(t),y(t);y(t+\tau)|z(t))=0$.

\begin{figure}
	\begin{center}
		\includegraphics[scale=0.25]{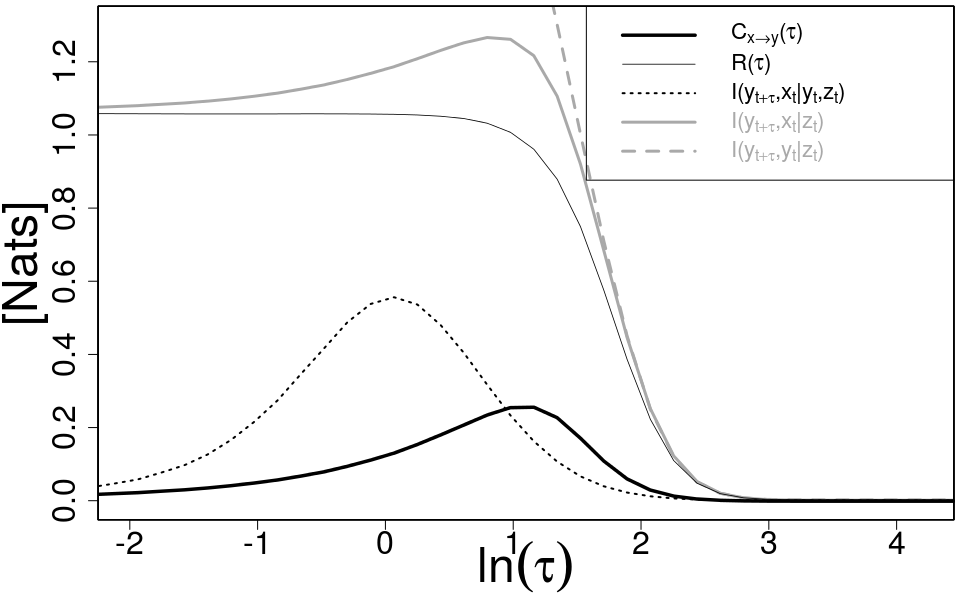}
		\caption{\label{sette} Feed-forward loop, the 3-dimensional general case. Causal influence $x\longrightarrow y$ (numerical simulation). The parameters are $t_{rel}=10$, $\gamma=\alpha_x=\alpha_y=1$, $\beta_x=\beta_y=0.2$, $D_z=10$, $D_x=D_y=0.1$.}
	\end{center}
\end{figure}

When $\gamma\neq0$ our causal influence measure $C_{x\rightarrow y}(\tau)$ would detect the presence of the $x\rightarrow y$ influence (numerical results in fig.\ref{sette}). We verified numerically that the causal influence is correctly zero for the cases $C_{y\rightarrow x}=C_{x\rightarrow z}=C_{y\rightarrow z}=0$. The transfer entropy $I(y(t+\tau),x(t)|y(t),z(t))$ goes to $0$ for $\tau \rightarrow 0$ because the white noise $\sqrt{D_y}\,\Gamma_y$ becomes dominant.

We note that even without a direct interaction $z\rightarrow y$, that is $\alpha_y=0$, the causal influence $C_{z\rightarrow y}$ can be positive due to the indirect influence $z\rightarrow x\rightarrow y$. The bigger is the number of indirect passages between the considered nodes, the longer is the time period $\tau$ after which the causal influence is seen.

\section{Discussion}

Within the study of information flows in linear response models we built up a quantitative definition of causal influence. The causal influence is defined as the unique information\cite{barrett2015exploration} on the evolution of the response $y(t+\tau)$ given by the signal $x(t)$, i.e. the difference between the mutual information $I(x(t),y(t+\tau))$ and the redundancy $R(x(t),y(t);y(t+\tau))$. This is based on the measure of redundant information $R$ that we define as a composition of the mutual information between the two sources $I(x(t),y(t))$ and the total information that they give together on the output $I((x(t),y(t)),y(t+\tau))$. The formula for this composition (eq.\ref{R}) we did not derive from first principles but is a choice inspired on an analogy with Markov chains.
We stress that the causal influence is a function of the mutual information, of the time-lagged mutual information and of the Transfer Entropy. With our definition of redundancy the causal influence results in a peak function of time starting from $0$ at $\tau=0$, meaning that the effects of causality are seen gradually over time, reflecting our view of the effects being visible only after the causes.
Importantly, when there is no influence of a variable $x$ on the dynamics of another variable $y$, i.e. in the network of direct influences there's no directed path starting from $x$ and arriving in $y$, the causal influence is correctly zero even if the two variables are highly correlated.

The main difference of our information decomposition to the one of Williams and Beer is that our redundancy $R$ is explicitly dependent on the information shared between the two variables giving the redundant information on the third one and is always less or equal to that. Taking as redundancy just the minimum mutual information leads to threshold effects that do not seem to be appropriate to describe linear response models.

Finally we try to understand whether the concept of causal influence still makes sense in systems with feedbacks. When the variable $x$ is influencing $y$ and vice versa forming a feedback loop, we can't define anymore a signal and a response. The $x(t)$ at time $t$ is influencing the evolution of the response $y(t+\tau)$ at time $t+\tau$ in many ways: directly and also indirectly through the loop $x(t)\rightarrow y(t+t')\rightarrow x(t+t'')\rightarrow y(t+\tau)$ with $\tau>t''>t'>0$, but also through the loops $x(t)\rightarrow y(t+t')\rightarrow x(t+t'')\rightarrow y(t+t''')\rightarrow x(t+t'''')......\rightarrow y(t+\tau)$ and so on.
These "successive" influences are in opposing directions for negative feedback loops and this implies the information measures to oscillate over time and our measure of causal influence to oscillate as well and to assume also negative values. Since the mutual information $I(x(t),y(t+\tau))$ is periodically assuming $0$, we would have oscillating causal influence with any definition of $R$ and we may conclude that the point-to-point communication scheme $(t,t+\tau )$ is not appropriate for a definition of causal influence in the presence of feedbacks.

An additional problem in defining a measure of causal influence is the information decomposition itself. What is the meaning of decomposing entropies? The mutual overlap between the two probability distributions $P(y(t+\tau)|y(t))$ and $P(y(t+\tau)|x(t))$ could be defined as $\braket{\int_{-\infty}^{\infty}P(y(t+\tau)|y(t))P(y(t+\tau)|x(t))dy(t+\tau)}_{x(t),y(t)}$ but then is not easy to say if such quantity can be used to define a measure of overlap (redundancy) in the space of Shannon entropies.  

In general, the causal influence quantifies the effective strength of asymmetric causal interactions and the time scale over which the effects are seen. Our measure is a good description of the dynamics of influences in linear response models. However, a generalization to nonlinear and non-Gaussian dynamics and for systems that have feedbacks is needed to approach the data analysis in complex systems.

\begin{widetext}
\section*{Appendix A: THE CORRELATION DUE TO A COMMON PARENT RESULTS IN ZERO CAUSAL INFLUENCE}
We study the particular case of the system of eq.\ref{trio} without influence of the $x$ on the $y$, i.e. with $\gamma=0$. We calculate here all the information measures needed to show that the causal influence $C_{x\rightarrow y}(\tau)$ is zero. We start from those quantities which are found also in the BLRM. The conditional expectation values and the standard deviations for the couples $zx$ and $zy$ are symmetric so we write them just once:
\begin{eqnarray}\label{A1}
&\braket{z(t-\tau+t')z(t-\tau+t'+t'')|z(t)}=\nonumber\\&=\int_{-\infty}^{+\infty}P(z(t-\tau+t'+t'')=\xi|z(t))\xi \braket{z(t-\tau+t')|z(t-\tau+t'+t'')=\xi}d\xi= \nonumber\\ 
&=e^{-\frac{t''}{t_{rel}}}(z^2(t) e^{-\frac{2(\tau-t'-t'')}{t_{rel}}}+\sigma^2_z (1-e^{-\frac{2(\tau-t'-t'')}{t_{rel}}}))
\end{eqnarray}
\begin{eqnarray}\label{A2}
&\braket{z(t-\tau+t')x(t-\tau+t')|z(t)}=\nonumber\\&=\int_{-\infty}^{+\infty}P(z(t-\tau+t')=\xi|z(t))\xi \braket{x(t-\tau+t')|z(t-\tau+t')=\xi}d\xi= \nonumber\\ 
&=\frac{\alpha_x t_{rel}}{\beta_x t_{rel}+1}(z^2(t) e^{-\frac{2(\tau-t')}{t_{rel}}}+\sigma^2_z (1-e^{-\frac{2(\tau-t')}{t_{rel}}}))
\end{eqnarray}
\begin{eqnarray}\label{A3}
&\braket{z(t-\tau+t')x(t)|z(t)}=\nonumber\\
&=\braket{z(t-\tau+t')x(t-\tau+t')|z(t)} e^{-\beta_x (\tau-t')}+ \alpha_x \int_{0}^{\tau-t'}\braket{z(t-\tau+t')z(t-\tau+t'+t'')|z(t)}e^{-\beta_x (\tau-t'-t'')} dt''=  \nonumber\\
&=\sigma^2_z  \frac{2 \alpha_x  t_{rel}}{\beta^2_x t_{rel}^2-1}(e^{-\frac{\tau-t'}{t_{rel}}}-e^{-\beta_x (\tau-t')})
+z^2(t) \frac{\alpha_x t_{rel}}{\beta_x t_{rel} +1} e^{-\frac{\tau-t'}{t_{rel}}}
\end{eqnarray}
\begin{eqnarray}\label{A4}
\sigma^2_y=\sigma^2_z \frac{\alpha_y^2 t_{rel}}{\beta_y(1+\beta_y t_{rel})} +\frac{D_y}{2 \beta_y}
\end{eqnarray}
\begin{eqnarray}\label{A5}
\sigma^2_{y(t+\tau)|z(t)}=\sigma_y^2 -(\frac{\sigma_z \alpha_y t_{rel}}{\beta_y t_{rel} -1})^2 (e^{-\frac{\tau}{t_{rel}}}-\frac{2 e^{-\beta_y \tau}}{\beta_y t_{rel} +1})^2
\end{eqnarray}
\begin{eqnarray}\label{A6}
&\sigma^2_{y(t+\tau)|x(t),y(t),z(t)}=\sigma^2_{y(t+\tau)|y(t),z(t)}= \nonumber\\
&=\frac{\sigma_z^2 \alpha_y^2}{\beta_y (\beta_y+1/t_{rel})(\beta_y-1/t_{rel})^2} [(1-\beta_y t_{rel})^2-e^{-2\beta_y \tau}(1+\beta_y t_{rel})+e^{-(\beta_y +\frac{1}{t_{rel}})\tau}4\beta_y t_{rel}-e^{-\frac{2\tau}{t_{rel}}}\beta_y t_{rel}(1+\beta_y t_{rel})]+\nonumber\\& + \frac{D_y}{2 \beta_y} (1-e^{-2 \beta_y \tau}) 
\end{eqnarray}
\begin{eqnarray}\label{A7}
\braket{y(t-\tau)x(t)|z(t)}=\braket{y(t)x(t)|z(t)} e^{\beta_y \tau} -\alpha_y \int_{0}^{\tau}\braket{z(t-\tau+t')x(t)|z(t)} e^{\beta_y t'} dt'
\end{eqnarray}
$\braket{y(t-\tau)x(t)|z(t)}\rightarrow 0$ for $\tau \rightarrow \infty$ because the knowledge of $x(t)$ gives asymptotically no information on the distant past of $y$ (even with the condition $z(t)$), then:
\begin{eqnarray}\label{A8}
\braket{y(t)x(t)|z(t)}=\frac{\alpha_x \alpha_y t_{rel}^2}{(\beta_x t_{rel}+1)(\beta_y t_{rel}+1)}(z^2(t)+\frac{2 \sigma^2_z}{t_{rel}(\beta_x+\beta_y)})
\end{eqnarray}
Since $\braket{z(t+t')x(t)|z(t)}=\braket{z(t+t')|z(t)}\braket{x(t)|z(t)}$, whose quantities we know from eq.\ref{ecinque}-\ref{esei}, we can easily calculate $\braket{y(t+\tau)x(t)|z(t)}=\braket{y(t)x(t)|z(t)} e^{-\beta_y \tau} +\alpha_y \int_{0}^{\tau}\braket{z(t+t')x(t)|z(t)} e^{-\beta_y (\tau-t')} dt'$ and then:
\begin{eqnarray}\label{A9}
&\braket{y(t+\tau)x(t)|z(t)}-\braket{y(t+\tau)|z(t)}\braket{x(t)|z(t)}=\nonumber \\
&=\sigma^2_z \frac{2 \alpha_x \alpha_y t_{rel} e^{-\beta_y \tau}}{(\beta_x t_{rel}+1)(\beta_y t_{rel}+1)(\beta_x+\beta_y)}
\end{eqnarray}
which is independent of the condition $z(t)$, as is typically the case for linear systems.
The information measures are easily calculated in the Gaussian case:
\begin{eqnarray}\label{A10}
I_{tot}=\ln(\frac{\sigma^2_{y(t+\tau)|z(t)}}{\sigma^2_{y(t+\tau)|x(t),y(t),z(t)}})
\end{eqnarray}
\begin{eqnarray}\label{A11}
C(x(t),y(t+\tau)|z(t))=\frac{\braket{y(t+\tau)x(t)|z(t)}-\braket{y(t+\tau)|z(t)}\braket{x(t)|z(t)}}{\sigma_{y(t+\tau)|z(t)}\sigma_{x(t)|z(t)}}
\end{eqnarray}
\begin{eqnarray}\label{A12}
I(x(t),y(t+\tau)|z(t))=-\frac{1}{2} \ln(1-C^2(x(t),y(t+\tau)|z(t)))
\end{eqnarray}
Using the definition of redundancy (eq.\ref{R}), $R(\tau)=\frac{1}{2} \ln(\frac{e^{2(I_{xy}+I_{tot})}}{e^{2I_{xy}}+e^{2I_{tot}}-1})$, with $I_{xy}=-\frac{1}{2} \ln(1-C^2(x(t),y(t)|z(t)))$ we obtain the expected result:
\begin{eqnarray}\label{A13}
C_{x\rightarrow y}(\tau)=I(x(t),y(t+\tau)|z(t))-R(x(t),y(t);y(t+\tau)|z(t))=0
\end{eqnarray}
\end{widetext}

\section*{Acknowledgments}
We thank Dr. Marco Scazzocchio for helpful discussions. This work has been funded by the DFG (Graduiertenkolleg 1772 for Computational Systems Biology) and the BMBF.

\bibliography{Causal_influence}


\end{document}